# Reliable LC-MS/MS-based method for trace level determination of 50 medium to highly polar pesticide residues in sediments and ecological risk assessment.

Maria Vittoria Barbieri [a], Cristina Postigo [a*], Simon Monllor-Alcaraz [a], Damià Barceló [a,b], Miren López de Alda [a*]

[a] *Water, Environmental and Food Chemistry Unit (ENFOCHEM), Department of Environmental Chemistry, Institute of Environmental Assessment and Water Research (IDAEA-CSIC), C/Jordi Girona 18-26, 08034 Barcelona, Spain*

[b] *Catalan Institute for Water Research (ICRA), Emili Grahit, 101, Edifici H2O, Parc Científic i Tecnològic de la Universitat de Girona, 17003 Girona, Spain.*

***Corresponding Authors**: Cristina Postigo (0000-0002-7344-7044)
Miren López de Alda (0000-0002-9347-2765)
Institute of Environmental Assessment and Water Research (IDAEA-CSIC)
Department of Environmental Chemistry
C/ Jordi Girona 18-26, 08034 Barcelona, Spain.
cprqam@cid.csic.es, mlaqam@cid.csic.es
Tel: +34-934-006-100, Fax: +34-932-045-904

## ABSTRACT

The occurrence of polar pesticides in sediments has not been extensively investigated because of their relatively poor hydrophobicity and apparently less persistence in the environment. However, their continuous release into the aquatic systems calls for the evaluation of their potential accumulation in sediments and the role of this matrix as a potential source of these compounds. Considering this, a method based on pressurized liquid extraction (PLE), extract clean-up by solid phase extraction (SPE) and analyte determination by liquid chromatography coupled to tandem mass spectrometry (LC-MS/MS) was developed and validated to analyze 50 relevant (frequently used and/or regulated or found in water) medium to highly polar pesticides in sediments. The method showed good performance regarding accuracy (relative recoveries between 76% and 124%), precision (relative standard deviation values <20%), sensitivity (LODs in the low ng/g for most compounds), linearity (coefficients of determination >0.99), and matrix effects (negligible for all analytes). The use of an isotope dilution approach for quantification ensures results reliability. As a part of the validation process, the method was applied to the analysis of the target pesticides in sediments from the Llobregat River (NE Spain) showing the presence of five of them, namely, terbutryn, dichlorvos, terbuthylazine, diazinon, and irgarol. All 5 pesticides, due to both the concentrations found and their physical-chemical characteristics, demonstrate high potential for bioaccumulation and risk to the aquatic organisms. Additional multidisciplinary studies that investigate pesticides occurrence in the different aquatic compartments and evaluate the potential risks for aquatic ecosystems are required to assess the environmental impact and significance of the presence of pesticides in sediments.

## 1. Introduction

Pollutants may get adsorbed onto suspended particulate matter present in surface water. These particles may eventually deposit on the bottom and at the banks of water bodies, forming sediment layers. Thus, sediments play a fundamental role in the transport and fate of pollutants [1]. As a pollution sink, sediments also reflect long-term water pollution and can be used to evaluate the impact of human activities because changes in their pollution pattern are quite slow [2]. Pesticides are among those organic pollutants that may end up in freshwater sediments. However, their presence in sediments has been little studied and, when measured in aquatic systems, studies were focused in the freely dissolved phase.

Most of the few works that studied their fate in sediments focus on organochlorine pesticides [3-6], because of their low water solubility, high potential to adsorb on particles, persistence and toxicity. However, in the late 1970s most organochlorine pesticides started to be banned in developing countries and consequently many other pesticides started to be synthesized and introduced in the market and are nowadays commonly used. In this regard, polar pesticides were presented as an attractive alternative to organochlorine pesticides, because of their high water solubility, low organic carbon sorption coefficient ($K_{oc}$), and low environmental persistence. Nevertheless, their large use in agriculture, the main source of pesticide pollution in the environment [7], has led to their ubiquitous presence in water ecosystems. This makes the screening of these pesticides in sediments extremely important for understanding their environmental fate and assessing the potential risks that they may pose for aquatic organisms [2].

Several pesticides are included in the list of priority substances that determine the achievement, or not, of a good chemical status of surface water bodies in Europe. Consequently, their concentrations cannot exceed certain Environmental Quality Standards (EQS) established in surface waters, and in some cases also in biota [8]. As for now, EQS have not been set yet for sediments, but the Directive establishes that this can be done at EU member state level and, in any case, long-term trend monitoring of priority substances concentrations in sediments has to be performed in order to prevent deterioration of surface water bodies. Therefore, the development of high sensitivity and reliable analytical methods to measure low levels of this type of organic pollutants in such a complex matrix is required for this purpose.

In the last 20 years, liquid chromatography coupled to tandem mass spectrometry (LC-MS/MS) has been recognized as the technique of choice for the determination of polar compounds and their transformation products in environmental matrices [9, 10] and food [11, 12] because it provides good separation for ionic and small polar analytes (which includes most metabolites) and very good performance in terms of sensitivity and selectivity when operated in the selected reaction monitoring (SRM) mode. In the last decade, this technique has been selected to analyze these organic substances in sediments [13-18]. For this, a wide range of extraction procedures have been used: Soxhlet extraction [19], matrix solid-phase dispersion (MSPD) extraction [15], QuEChERS (quick, easy, cheap, effective, rugged and safe) approaches [18, 20, 21], ultrasonic solvent extraction [22, 14], microwave-assisted extraction (MAE) [17], and pressurized liquid extraction (PLE) [23, 18, 24]. The extraction of organic chemicals from sediments is always a critical step, due to the strong interactions, e.g., through ion exchange or covalent and H-bonding, that may occur between them and the organic matter and mineral surface of the sediment [25]. The precision of the extraction step can be improved with the use of automated techniques such as PLE. Besides, additional advantages that PLE presents over other extraction methods are the reduction of the extraction time and the requirement of low solvent volumes.

Thus, the main objective of this work was to develop and validate an analytical methodology based on PLE, solid phase extraction (SPE) purification, and LC-MS/MS analysis, for the simultaneous determination of 50 medium to highly polar pesticides in sediments. The list of compounds, which includes 40 pesticides and 10 transformation products, was designed taking into account three main criteria: i) pesticide use at European level, ii) environmental regulations, i.e., pesticides included in the list of priority substances in water [8] or in the European watch list [26], and iii) possibility of analysis by means of LC-MS/MS. The investigated pesticides are used for different purposes (herbicides, fungicides, biocides) and applied in different sectors (agricultural, urban or industrial uses). They belong to different chemical classes (organophosphates, phenylureas, chloroacetamides, neonicotinoids, triazines, acidic pesticides, and other classes), and therefore present a wide range of physical-chemical properties ($K_{oc}$, $K_{ow}$, solubility, GUS index, etc.). Thus, a comprehensive study of their fate and impact in aquatic systems subject to different environmental stressors requires the development and application of multi-residue methods allowing their simultaneous analysis in the various environmental compartments, including sediments. Moreover, to the authors' knowledge, some of the target pesticides, namely, azinphos-methyl oxon, , fenitrothion oxon, and malaoxon, were not previously investigated in sediment samples.

Finally, this method was applied to determine the occurrence of the selected pesticides in sediments of the Llobregat River basin (Catalonia, Spain). The concentrations observed were used to provide a first picture of the chemical status of this freshwater ecosystem and identify the most relevant threats so that appropriate mitigation measurements can be adopted.

## 2. Materials and methods

### *2.1 Chemicals*

High purity standards (96-99.9%) of the 50 pesticides and 44 isotopically-labeled compounds used as surrogate standards were purchased from Fluka (Sigma-Aldrich, Steinheim, Germany) or Dr. Ehrenstorfer (LGC Standards, Teddington, UK). Table 1 lists the target analytes including their chemical class, relevant physical-chemical properties (solubility in water, $K_{oc}$, $K_{ow}$, GUS index, and others) and actual legislative status regarding their use (approved or not approved). Stock individual standard solutions were prepared in methanol (MeOH), except in the case of simazine that was prepared in dimethyl sulfoxide, at a concentration of 1000 µg/mL, and stored in amber glass bottles in the dark at -20 °C. A mixture containing the surrogate standards at a concentration of 1000 ng/mL and working standard solutions containing the target pesticides at diverse concentrations (from 0.01 to 1000 ng/mL) were prepared by dilution of the stock individual solutions in MeOH. Pesticide grade-solvents MeOH, acetone (ACE), acetonitrile (ACN), dichloromethane (DCM), formic acid (FA) and LC grade-water were supplied by Merck (Darmstadt, Germany). Ottawa Sand was purchased from Applied Separations (Allentown, PA) and alumina from Merck (Darmstadt, Germany).

**Table 1.** Target pesticides and main physical-chemical properties.

| Analyte | Chemical class | Formula‡ | Legislative status‡ | MM (g mol$^{-1}$)‡ | Solubility (mg L$^{-1}$) ‡ | $K_{oc}$ (mL g$^{-1}$) ‡ | $K_{ow}$ logP‡ | Henry´s (Pa m$^3$ mol$^{-1}$) ‡ | GUS‡ | DT50‡ | Pka‡ |
|---|---|---|---|---|---|---|---|---|---|---|---|
| **2,4-D** | Alkylchlorophenoxy | $C_8H_6Cl_2O_3$ | ✓ | 221.04 | 24300 | 39 | -0.82 | 4.0 X $10^{-06}$ | 1.69 | 4.4 | 3.40 |
| **Acetamiprid** | Neonicotinoid | $C_{10}H_{11}ClN_4$ | ✓ | 222.67 | 2950 | 200 | 0.80 | 5.3 X $10^{-08}$ | 0.40 | 1.6 | 0.7 |
| **Alachlor** | Chloroacetamide | $C_{14}H_{20}ClNO_2$ | X | 269.77 | 240 | 335 | 3.09 | 3.2 X $10^{-03}$ | 1.08 | 14 | 0.62 |
| **Atrazine** | Triazine | $C_8H_{14}ClN_5$ | X | 215.68 | 35 | 100 | 2.70 | 1.5 X $10^{-04}$ | 3.2 | 75 | 1.7 |
| **Azinphos-ethyl** | Organophosphate | $C_{12}H_{16}N_3O_3PS_2$ | X | 345.38 | 4.5 | 1500 | 3.18 | 3.1 X $10^{-06}$ | 1.4 | 50 | n/a |
| **Azinphos-methyl** | Organophosphate | $C_{10}H_{12}N_3O_3PS_2$ | X | 317.32 | 28 | 1112 | 2.96 | 5.7 X $10^{-06}$ | 1.42 | 10 | 5 |
| **Azinphos-methyl-oxon** | Metabolite | $C_{10}H_{12}N_3O_4PS$ | - | 301.26 | 2604* | 10 * | 0.77* | 6.2 x $10^{-13}$ * | - | - | n/a |
| **Bentazone** | Benzothiazinone | $C_{10}H_{12}N_2O_3S$ | ✓ | 240.30 | 7112 | 55 | -0.46 | 7.2 X $10^{-05}$ | 2.89 | 20 | 3.51 |
| **Bromoxinil** | Hydroxybenzonitrile | $Br_2C_6H_2(OH)CN$ | ✓ | 276.90 | 38000 | 302 | 0.27 | 8.7 X $10^{-07}$ | 0.03 | 1.04 | 3.86 |
| **Chlorfenvinphos** | Organophosphate | $C_{12}H_{14}Cl_3O_4P$ | X | 359.60 | 145 | 680 | 3.80 | 1.4 X $10^{-01}$ * | 1.83 | 40 | n/a |
| **Chlortoluron** | Phenylurea | $C_{10}H_{13}ClN_2O$ | ✓ | 212.68 | 74 | 196 | 2.50 | 1.4 X $10^{-05}$ | 3.02 | 45 | n/a |
| **Cyanazine** | Triazine | $C_9H_{13}ClN_6$ | X | 240.69 | 171 | 190 | 2.10 | 6.6 X $10^{-06}$ | 2.07 | 16 | 12.9 |
| **Clothianidin** | Neonicotinoid | $C_6H_8ClN_5O_2S$ | ✓ | 249.68 | 340 | 123 | 0.90 | 2.9 X $10^{-11}$ | 4.91 | 545 | 11.1 |
| **Deisopropylatrazine** | Metabolite | $C_5H_8ClN_5$ | - | 173.60 | 980 | 130 | 1.15 | 980 | - | - | n/a |
| **Desethylatrazine** | Metabolite | $C_6H_{10}ClN_5$ | - | 187.63 | 2700 | 110 | 1.51 | 1.6 X $10^{-04}$ | 4.37 | 2.23^ | n/a |
| **Diazinon** | Organophosphate | $C_{12}H_{21}N_2O_3PS$ | X | 304.35 | 60 | 609 | 3.69 | 6.1 X $10^{-02}$ | 1.14 | 9.1 | 2.6 |
| **Dichlorvos** | Organophosphate | $C_4H_7Cl_2O_4P$ | X | 220.98 | 18000 | 50 | 1.90 | 2.6 X $10^{-02}$ | 0.69 | 2 | n/a |
| **Diflufenican** | Carboxamide | $C_{19}H_{11}F_5N_2O_2$ | ✓ | 394.29 | 0.05 | 5504 | 4.20 | 1.2 X $10^{-02}$ | 1.51 | 94.5 | n/a |
| **Dimethoate** | Organophosphate | $C_5H_{12}NO_3PS_2$ | ✓ | 229.26 | 25900 | 25* | 0.75 | 1.4 X $10^{-06}$ | 1.01 | 2.5 | n/a |
| **Diuron** | Phenylurea | $C_9H_{10}Cl_2N_2O$ | ✓ | 233.09 | 35.6 | 680 | 2.87 | 2.0 X $10^{-06}$ | 1.83 | 146.6 | n/a |
| **Fenitrothion** | Organophosphate | $C_9H_{12}NO_5PS$ | X | 277.23 | 19 | 2000 | 3.32 | 9.9 X $10^{-03}$ | 0.48 | 2.7 | n/a |
| **Fenitrothion oxon** | Metabolite | $C_9H_{12}NO_6P$ | - | 261.17* | 301 * | 21* | 1.69* | 4.0 X $10^{-1}$ * | - | - | n/a |
| **Fenthion** | Organophosphate | $C_{10}H_{15}O_3PS_2$ | X | 278.33 | 4.2 | 1500 | 4.84 | 2.4 X $10^{-02}$ | 1.26 | 22 | n/a |
| **Fenthion oxon** | Metabolite | $C_{10}H_{15}O_4PS$ | - | 262.26* | 213.5* | 57 * | 2.31* | 3.0x$10^{-9}$ * | - | - | n/a |
| **Fenthion oxon sulfone** | Metabolite | $C_{10}H_{15}O_6PS$ | - | 294.03* | 7602* | 13* | 0.28* | 2.4 x $10^{-11}$ * | - | - | n/a |
| **Fenthion oxon sulfoxide** | Metabolite | $C_{10}H_{15}O_5PS$ | - | 278.26* | 1222* | 11* | 0.15* | 9.5 x $10^{-8}$ * | - | - | n/a |
| **Fenthion sulfone** | Metabolite | $C_{10}H_{15}O_5PS_2$ | - | 310.33* | 190.4* | 235 | 2.05* | 1.1x$10^{-8}$ * | - | - | n/a |
| **Fenthion sulfoxide** | Metabolite | $C_{10}H_{15}O_4PS_2$ | - | 294.33* | 3.72* | 183 | 1.92* | 7.0x$10^{-6}$ * | - | - | n/a |
| **Fluroxypir** | Pyridine compound | $C_7H_5Cl_2FN_2O_3$ | ✓ | 255.03 | 6500 | 10* | 0.04 | 169 X $10^{-10}$ | 2.42 | 13.1 | 2.94 |
| **Imidacloprid** | Neonicotinoid | $C_9H_{10}ClN_5O_2$ | ✓ | 255.66 | 610 | 6719 | 0.57 | 1.7 X $10^{-10}$ | 3.74 | 191 | n/a |
| **Irgarol** | Triazine | $C_{11}H_{19}N_5S$ | X | 253.37 | 7 | 1569 | 3.95 | $1.3 \times 10^{-07}$* | - | - | n/a |
| **Isoproturon** | Phenyluera | $C_{12}H_{18}N_2O$ | X | 206.28 | 70.2 | 251* | 2.5 | $1.5 \times 10^{-05}$ | 2.07 | 12 | n/a |
| **Linuron** | Phenyluera | $C_9H_{10}Cl_2N_2O_2$ | X | 249.09 | 63.8 | 843 | 3 | $2.0 \times 10^{-04}$ | 2.21 | 57.6 | n/a |
| **Malaoxon** | Metabolite | $C_{10}H_{19}O_7PS$ | - | 314.29* | 7500* | 4650* | 0.52* | $1.2 \times 10^{-08}$ * | - | - | n/a |
| **Malathion** | Organophosphate | $C_{10}H_{19}O_6PS_2$ | ✓ | 330.36 | 148 | 1800 | 2.75 | $1.0 \times 10^{-03}$ | -1.28 | 0.17 | n/a |
| **MCPA** | Organophosphate | $C_9H_9ClO_3$ | ✓ | 200.62 | 29390 | 29* | -0.81 | $5.5 \times 10^{-05}$ | 2.94 | 24 | 3.73 |
| **Mecoprop** | Aryloxyalkanoic acid | $C_{10}H_{11}ClO_3$ | X | 214.65 | 250000 | 47 | -0.19 | $2.2 \times 10^{-04}$ | 2.29 | 8.2 | 3.11 |
| **Methiocarb** | Carbamate | $C_{11}H_{15}NO_2S$ | ✓ | 225.31 | 27 | 182* | 3.18 | $1.2 \times 10^{-04}$ | 0.55 | 2.94 | n/a |
| **Metolachlor** | Chloroacetamide | $C_{15}H_{22}ClNO_2$ | X | 283.80 | 530 | 120 | 3.40 | $2.4 \times 10^{-03}$ | 2.10 | 90 | n/a |
| **Molinate** | Thiocarbamate | $C_9H_{17}NOS$ | X | 187.30 | 1100 | 190 | 2.86 | $6.9 \times 10^{-01}$ | 2.49 | 28 | n/a |
| **Pendimethalin** | Dinitroaniline | $C_{13}H_{19}N_3O_4$ | ✓ | 281.31 | 0.33 | 17491 | 5.40 | $2.7 \times 10^{-03}$ | -0.32 | 182.3 | 2.8 |
| **Propanil** | Anilide | $C_9H_9Cl_2NO$ | X∞ | 218.08 | 95 | 149 | 2.29 | $4.4 \times 10^{-04}$ | -0.51 | 0.4 | 19.1 |
| **Quinoxyfen** | Quinoline | $C_{15}H_8Cl_2FNO$ | ✓ | 308.13 | 0.05 | 23° | 4.66 | $3.2 \times 10^{-02}$ | -0.93 | 308 | n/a |
| **Simazine** | Triazine | $C_7H_{12}ClN_5$ | X | 201.66 | 5 | 130 | 2.30 | $5.6 \times 10^{-05}$ | 2 | 60 | 1.62 |
| **Terbuthylazine** | Triazine | $C_9H_{16}ClN_5$ | ✓ | 229.71 | 6.6 | 329* | 3.40 | $3.2 \times 10^{-03}$ | 3.07 | 72 | 1.9 |
| **Terbutryn** | Triazine | $C_{10}H_{19}N_5S$ | X | 241.36 | 25 | 2432 | 3.66 | $1.5 \times 10^{-03}$ | 2.4 | 74 | 4.3 |
| **Thiacloprid** | Neonicotinoid | $C_{10}H_9ClN_4S$ | ✓ | 252.72 | 184 | 615° | 1.26 | $5.0 \times 10^{-10}$ | 0.14 | 15.5 | n/a |
| **Thiamethoxam** | Neonicotinoid | $C_8H_{10}ClN_5O_3S$ | ✓ | 291.71 | 4100 | 56 | -0.13 | $4.7 \times 10^{-10}$ | 4.69 | 50 | n/a |
| **Thifensulfuron methyl** | Sulfonylurea | $C_{12}H_{13}N_5O_6S_2$ | ✓ | 387.39 | 54.1 | 28 | -1.65 | $3.3 \times 10^{-08}$ | 0.44 | 1.39 | 4 |
| **Triallate** | Thiocarbamate | $C_{10}H_6Cl_3NOS$ | ✓ | 304.7 | 4.1 | 3034 | 4.06 | 0.89 | 0.69 | 82 | n/a |

ǂ The PPDB, Pesticide Properties Database. http://sitem.herts.ac.uk/aeru/footprint/index2.htm. - Lewis, K.A., Tzilivakis, J., Warner, D. and Green, A. (2016). An international database for pesticide risk assessments and management. Human and Ecological Risk Assessment: An International Journal, 22(4), 1050-1064.

*Data estimated using the US Environmental Protection Agency EPISuite™ http://www.Chemspider.com.

° Kegley, S.E., Hill, B.R., Orme S., Choi A.H., PAN Pesticide Database, Pesticide Action Network, North America (Oakland, CA, 2016), http://www.pesticideinfo.org.

∞ EU pesticides database. https://bit.ly/1oxd00K.

^ Calculated using the mathematical formula: GUS = log10 (half-life) x [4 - log10 (Koc)].

MM: molecular mass; Solubility: solubility in water at 20 °C ; $K_{oc}$: organic carbon partition coefficient; $K_{ow}$: octanol-water partition coefficient; Henry's: Henry's law constant at 25ºC; GUS: leaching potential index; DT50: soil degradation potential, expressed as half-life in days; Pka: dissociation constant at 25 °C; n/a: data not available

### 2.2 *Extraction and clean-up of the sediment samples*

For sample extraction, a 22 mL stainless steel extraction cell was prepared by placing sequentially at the bottom of the cell one cellulose filter (0.45 µm pore size), 1 g of sand, and 1 g of alumina, previously activated by heating it at 80 °C for at least 24h (see Fig. 1). Then, an aliquot of 5 g of lyophilized and sieved (125 µm) sediment sample was weighed directly into the cell, spiked with the surrogate standard mixture at a concentration of 50 ng/g and left overnight in a fume hood for 12h to allow methanol evaporation and interaction of the surrogate standards with the matrix. The day after, 6 g of activated alumina was placed into the cell and mixed with the sediment to reduce the unintended extraction of matrix components other than the analytes of interest. Then, the cell was filled up to the top with sand, closed, and positioned in the PLE system, a Dionex Accelerated Solvent Extraction (ASE) 350 apparatus (Vertex Technics S.L., Barcelona). PLE extraction was performed using an acidified mixture of ACE and DCM (1:1, and 1% FA, v/v) as extracting solvent. The extraction conditions were as follows: pressure, 1600 psi; temperature, 100 °C; heating time, 5 min; purge time, 90 s; static time, 5 min. Two extraction cycles were performed, resulting in a total extraction time of 20 min per cell. Upon extraction, the final extract (≈ 20 mL), to which one mL of MeOH was added, was evaporated under nitrogen to 1 mL using a Biotage TurboVap® LV workstation (Vertex Technics S.L., Barcelona). Then, the one-mL extract was dissolved in 19 mL of LC-grade water for SPE clean-up. This was performed using a generic-purpose Oasis HLB sorbent (500 mg, 6 cc cartridges, Waters, Milford, MA, USA). Before loading the aqueous extract, the SPE sorbent was conditioned with 5 mL of MeOH:DCM (1:1) and 5 mL of LC-grade water. After extract loading, the sorbent was washed with 5 mL of LC-grade water to remove matrix interferences. Then, the sorbent was dried for 30 minutes, and analytes were eluted using 4 mL of MeOH:DCM (1:1). The eluate was evaporated under nitrogen to approximately 1.5 mL and finally reconstituted with MeOH to 5 mL before its transfer to 2-mL vials for LC-MS/MS analysis.

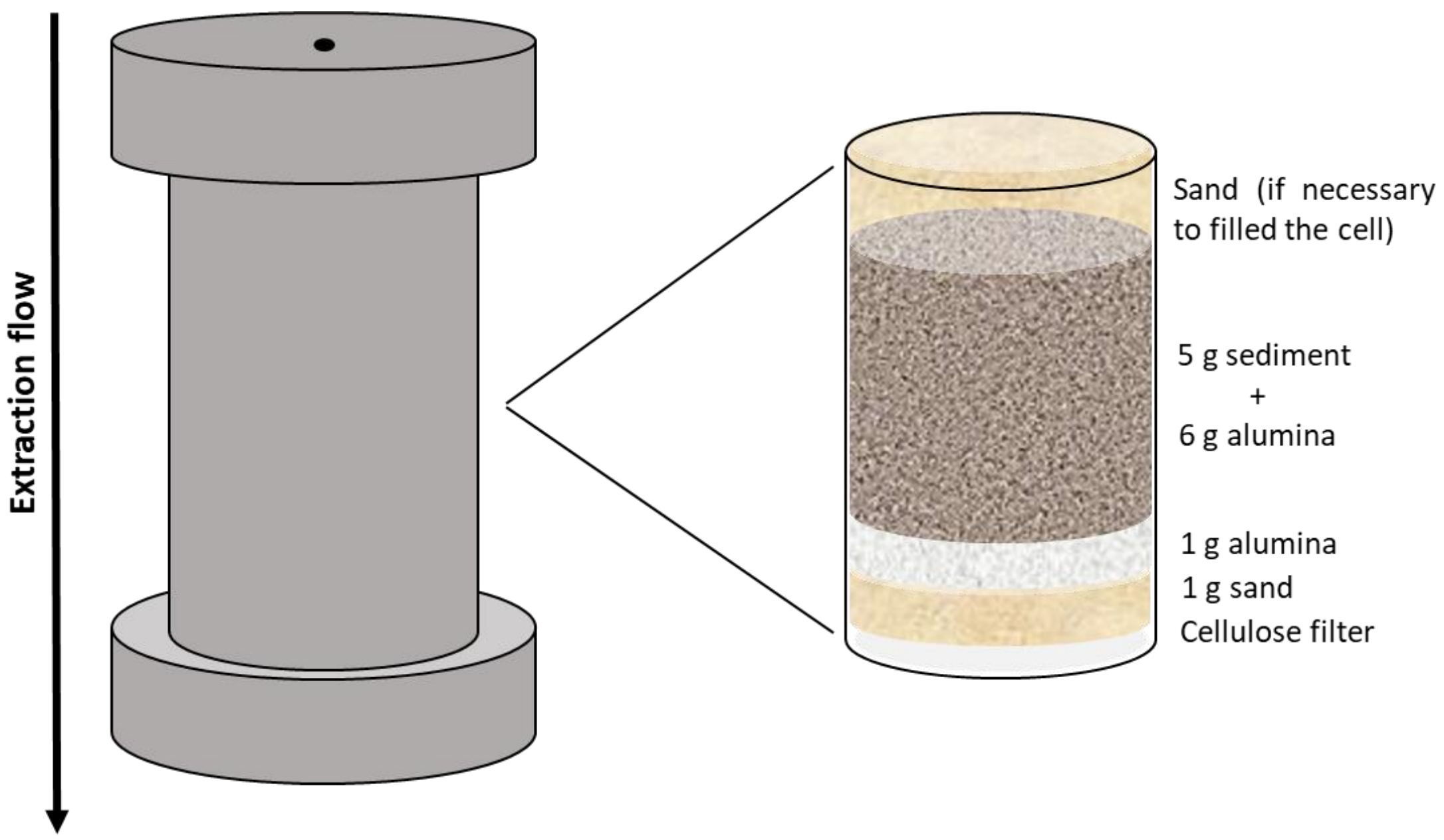

**Figure 1.** Scheme of the PLE extraction cell.

### *2.3 LC-HESI-MS/MS analysis*

Chromatographic separation was performed by means of an Aria$^{TM}$ LC system equipped with two Transcend quaternary pumps (Thermo-Fisher Scientific Inc.), using a Purospher STAR RP-18e column (150 × 2.1 mm, 2 µm particle diameter) (Merck, Darmstadt, Germany) and a mobile phase consisting of ACN and water at a flow rate of 0.2 mL/min. The volume of extract injected was 10 µL. The organic gradient used was as follows: 10% at time (t)= 0, 50% at t=2 min, 80% at t=12 min, 100% at t=13 min and maintained until t=16 min, and returned to 10 % in t=17 min. Initial conditions were held for 8 min (until t=25 min) for column re-equilibration.

MS/MS detection was carried out using a TSQ Quantiva triple-quadrupole mass spectrometer (Thermo Fisher Scientific Inc.) equipped with a heated electrospray ionization source (HESI). All analytes were determined in a single analytical run, being 43 compounds ionized in the positive (PI) mode and 7 compounds in the negative (NI) mode in one single acquisition window. The mass spectrometer was operated in the selected reaction monitoring (SRM) mode, acquiring two SRM transitions per target compound and one SRM transition per surrogate compound. MS acquisition conditions were as follows: ion spray voltage, 3500 V for PI and -2500 V for NI; vaporizer temperature, 280 °C; ion transfer tube temperature, 350 °C. Nitrogen was used as sheath gas, auxiliary gas, and sweep gas, and argon was used as collision gas, at a pressure of 2.5 mTorr. Thermo Xcalibur 3.0.63 software (Thermo Fisher Scientific Inc.) was used for instrument control, data acquisition, and evaluation.

### *2.4 Method performance*

The sediment sample used in the validation study was collected from the Guadalquivir River basin (South of Spain). Method validation was performed in terms of linearity, accuracy (recovery), precision (repeatability), sensitivity, and matrix effects.

Method linearity was evaluated in methanolic solutions containing the analytes within the concentration range 0.01 - 1000 ng/mL (equivalent to 0.01 ng/g d.w. and 1000 ng/g d.w., respectively, in sediment) and the surrogate standards at 50 ng/mL (50 ng/g in sediment). For this, eleven-point calibration curves using an isotope dilution approach were constructed. Linearity was expressed as the goodness of fit, i.e., the coefficient of determination ($R^2$), of the calibration data to a least-squares linear regression model, obtained using $1/x^2$ as a weighting factor.

Method accuracy was appraised from analyte absolute and relative recoveries obtained after n=6 replicate analysis of fortified sediment samples at three different concentration levels (low: 10 ng/g d.w., medium: 50 ng/g d.w., and high: 100 ng/g d.w.). The absolute recoveries were evaluated by comparing the analyte peak areas obtained in fortified sediment samples and methanolic solutions at equivalent concentrations. The relative recoveries were calculated by comparing the absolute recoveries of the analytes with those of their surrogates. Method precision was assessed from the repeatability of n=6 replicate analysis of fortified sediment samples at the aforementioned concentration levels and expressed as the relative standard deviation (RSD) of the response obtained.

Appraisal of method sensitivity was done through the analytes limit of detection (LOD) and quantification (LOQ). These values were estimated from the analysis of sediment samples fortified at the lowest level (10 ng/g d.w.) as the concentration of analyte that provides a signal-to-noise ratio of 3, in the case of LOD, and 10, in the case of LOQ. Method sensitivity was also assessed through the analyte limit of determination (LODet), i.e., the minimum concentration at which the analyte can be quantified using the first SRM transition (LOQ of SRM1) and confirmed with the second SRM transition (LOD of SRM2).

Evaluation of the effects that the matrix components had on analyte ionization, i.e., matrix effects (ME), was done by comparing the analyte peak areas obtained in n=3 sediment samples fortified with the target analytes after their extraction ($A_{matrix}$) and in a methanolic solution ($A_{std}$) at equivalent concentrations (10 ng/mL). Ionization suppression effects were associated to negative ME values (smaller peak areas in $A_{matrix}$ than in $A_{std}$), whereas ionization enhancement effects were associated to positive ME values (larger peak areas in $A_{matrix}$ than in $A_{std}$).

Background concentrations of target analytes in the sediment sample used in the validation study were considered, if present, in all calculations done to evaluate method performance.

### *2.5 Study site and sample collection*

The method was applied to evaluate the presence of the target pesticides in river sediments collected from the lower Llobregat River basin (Catalonia, Spain), one of the most important drinking water resources for the city of Barcelona and its metropolitan area (over 3 million people). The Llobregat River is a typical Mediterranean river: it presents flow fluctuations related to climate conditions and seasons, with drought periods, when dilution capacity of the river is decreased and consequently, the risk of contamination is increased [27], and peak rainfall events in spring (March-June) and autumn (September-December) that remove the river bottom. The intensive urban and industrial activities and surface runoff from agriculture also contribute to the flow variation and pollution of the river [28].

A sampling campaign was carried out in February 2017. Seven different selected sampling sites were distributed along the basin (to depict a downstream contamination profile) and land and concrete water channels adjacent to the main river (to assess the impact of agriculture), as shown in Fig. 2. Sediment samples were taken using a Van Veen drag, placed in an aluminum tray, and wrapped with aluminum foil, transported to the laboratory, and stored at -20 °C. Samples were lyophilized during 36 h with a LyoAlfa 6-50 freeze-dryer (Telstar), sieved through a 125 µm mesh for homogenization, and stored at -20 °C in the dark until their analysis.

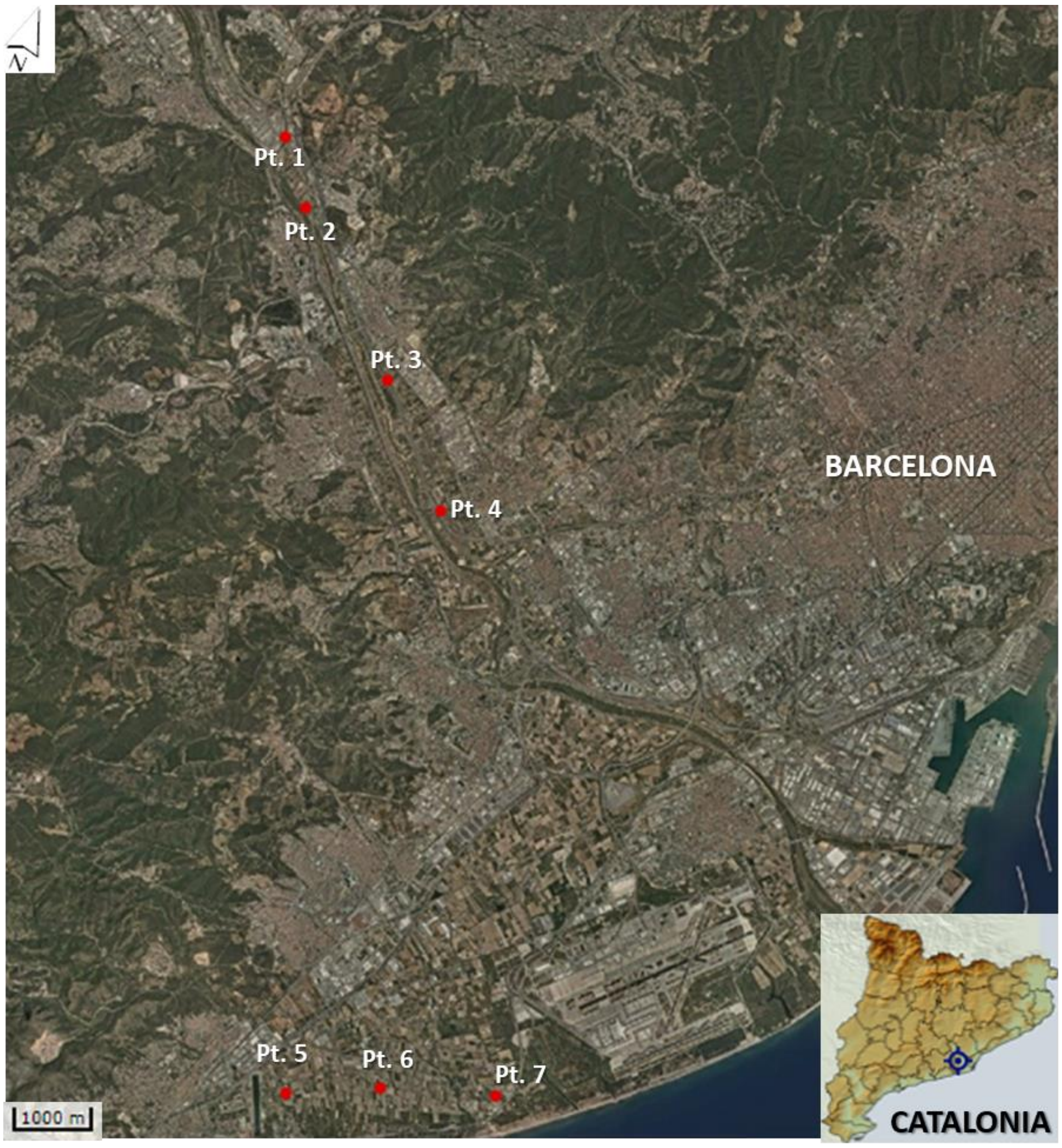

***Figure 2.*** *Map of the metropolitan area of Barcelona (Catalonia, Spain) with the Llobregat River basin and a detail of the locations of the sediment sampling points (Source: Generalitat de Catalunya, Visor ACA, http://sig.gencat.cat/visors/VISOR_ACA.html).*

## 3. Results and discussion

### *3.1 Method optimization*

The method developed was based on a previous analytical method established by the same research group for the analysis of 26 pesticides in sediments [29] that was improved for additional performance, automation, and time-saving. Main modifications include:

i) Expansion of the list of target analytes with 24 pesticides and 20 isotopically labeled analogs to cover relevant substances in future monitoring programs. The newly added pesticides include the EU priority substances dichlorvos, irgarol, quinoxyfen, and terbutryn [8], the pesticides included in the EU Watch List methiocarb, and the neonicotinoids acetamiprid, clothianidin, imidacloprid, thiacloprid, and thiamethoxam [26]; and banned substances like azinphos ethyl [30], azinphos-methyl [31] and one of its metabolites (azinphos-methyl oxon), and fenthion [32] and five of its metabolites (fenthion oxon, fenthion oxon sulfone, fenthion oxon sulfoxide, fenthion sulfone, fenthion sulfoxide). Moreover, pesticides currently used in Spain were also considered: bromoxynil [33], fluroxypyr [34], pendimethalin [35], thifensulfuron methyl [36], diflufenican [37].

ii) The use of the Dionex ASE 350 system instead of the PSE One system (Applied Separations, PA, USA), which allows increasing method automation. Contrary to the PSE One, in which only the extraction of one single cell could be programmed at a time, the Dionex ASE 350 allows the extraction of up to 24 different cells without intervention.

iii) The use of a mass spectrometer with a fast acquisition rate (TSQ Quantiva) instead of the TQD triple-quadrupole mass spectrometer from Waters. This allows reducing the acquisition time allocated for each SRM transition so that all transitions (144 in total) can be simultaneously acquired throughout the analytical run without losing sensitivity and reproducibility (obtaining sufficient points per peak). The optimum conditions for MS/MS determination of the compounds were individually optimized for each compound after injection of individual standard solutions in both PI and NI mode. The optimized conditions for the analysis of all analytes are provided in Tables 2 and 3.

iv) The use of a chromatographic column with a smaller particle size (2 µm vs. 5 µm), which improves peak resolution, and hence, separation efficiency, and allows reducing the analysis time from 40 min to 25 min. The composition of the LC mobile phase was maintained as it allows the analysis of all compounds in one single analytical run.

**Table 2.** MS/MS analysis of the target pesticides.

| Target pesticide | Precursor ion, m/z (RF Lens, V) | Product ion 1, m/z (CE, eV) | Product ion 2, m/z (CE, eV) | SRM1/ SRM2 |
|---|---|---|---|---|
| **Negative ionization mode (-)** | | | | |
| 2,4-D | 219 (35) | 162 (16) | 125 (28) | 32.2 |
| Bentazone | 239 (68) | 132 (28) | 117 (33) | 4.0 |
| Bromoxynil | 276 (82) | 81 (30) | 79 (30) | 1.1 |
| Fenitrothion | 262 (52) | 152 (21) | 122 (34) | 17.2 |
| Fluroxypyr | 255 (33) | 197 (16) | 235 (6) | 47.2 |
| MCPA | 199 (38) | 142 (17) | 105 (30) | 1.1 |
| Mecoprop | 213 (39) | 142 (17) | 140 (17) | 3.2 |
| **Positive ionization mode (+)** | | | | |
| Acetamiprid[a] | 223 (53) | 126 (22) | 90 (33) | 5.5 |
| Alachlor[b] | 270 (40) | 162 (21) | 132 (42) | 1.0 |
| Atrazine | 216 (58) | 174 (18) | 104 (28) | 3.0 |
| Azinphos-ethyl | 346 (37) | 137 (25) | 97 (32) | 1.6 |
| Azinphos methyl | 318 (30) | 132 (16) | 261 (7) | 1.6 |
| Azinphos methyl oxon | 324 (84) | 132 (22) | 148 (17) | 15.5 |
| Chlorfenvinphos[b] | 359 (60) | 170 (42) | 99 (27) | 1.6 |
| Chlortoluron | 213 (51) | 140 (25) | 104 (33) | 3.2 |
| Clothianidin[a] | 250 (43) | 169 (15) | 132 (18) | 1.2 |
| Cyanazine | 241 (59) | 214 (18) | 104 (30) | 4.1 |
| Desethylatrazine (DEA) | 188 (66) | 146 (18) | 104 (25) | 4.2 |
| Deisopropylatrazine (DIA) | 174 (58) | 104 (23) | 132 (18) | 1.2 |
| Diazinon | 305 (64) | 169 (22) | 153 (22) | 2.7 |
| Dichlorvos[b] | 221 (57) | 109 (14) | 145 (18) | 6.9 |
| Diflufenican | 395 (60) | 266 (24) | 246 (34) | 4.9 |
| Dimethoate | 230 (35) | 125 (22) | 157 (21) | 11.0 |
| Diuron[b] | 233 (51) | 160 (27) | 133 (41) | 1.2 |
| Fenitrothion oxon | 262 (66) | 216 (19) | 104 (22) | 1.5 |
| Fenthion | 279 (63) | 169 (20) | 247 (13) | 1.3 |
| Fenthion oxon | 263 (62) | 231 (16) | 216 (25) | 2.3 |
| Fenthion oxon sulfone | 295 (74) | 217 (20) | 91 (34) | 4.6 |
| Fenthion oxon sulfoxide | 279 (68) | 264 (20) | 262(22) | 6.2 |
| Fenthion sulfone | 311 (64) | 125 (21) | 233 (40) | 3.0 |
| Fenthion sulfoxide | 295 (68) | 280 (19) | 109 (33) | 1.5 |
| Imidacloprid[a] | 256 (51) | 209 (20) | 175 (20) | 1.6 |
| Irgarol[b] (Cybutryne) | 254 (57) | 198 (19) | 108 (30) | 10.1 |
| Isoproturon[b] | 207 (51) | 134 (23) | 91 (37) | 1.3 |
| Linuron | 249 (51) | 160 (19) | 133 (34) | 1.2 |
| Malaoxon (MOX) | 315 (48) | 99 (23) | 125 (33) | 13.6 |
| Malathion | 353 (70) | 227 (17) | 306 (15) | 4.3 |
| Methiocarb[a] | 266 (35) | 169 (9) | 121 (19) | 1.3 |
| Metolachlor | 284 (48) | 252 (16) | 176 (26) | 2.8 |
| Molinate | 188 (43) | 126 (13) | 98 (17) | 4.1 |
| Pendimethalin | 282 (33) | 212 (12) | 194 (19) | 7.4 |
| Propanil | 218 (53) | 127 (27) | 162 (16) | 1.5 |
| Quinoxyfen[b] | 310 (102) | 199 (34) | 216 (36) | 1.6 |
| Simazine[b] | 202 (61) | 132 (20) | 104 (25) | 1.0 |
| Terbuthylazine | 230 (52) | 174 (18) | 104 (32) | 6.1 |
| Terbutryn[b] | 242 (55) | 186 (19) | 158 (26) | 14.2 |
| Thiacloprid[a] | 253 (59) | 126 (23) | 90 (35) | 5.8 |
| Thiamethoxam[a] | 292 (47) | 132 (24) | 181 (23) | 1.1 |
| Thifensulfuron methyl | 388 (57) | 167 (18) | 205 (27) | 6.7 |
| Triallate | 304 (55) | 142 (40) | 83 (47) | 1.7 |

[a] Compound included in the EU watch list (2018/840/EC). [b] Compound included in the EU list of priority substances (2013/39/EC). CE: collision energy; SRM: selected reaction monitoring; SRM1/SRM2: peak area ratio.

**Table 3.** MS/MS analysis of the surrogate standards (isotopically labeled analogs were used for all analytes except 6).

| Surrogate standards | Precursor ion, m/z (RF Lens, V) | Product ion, m/z (CE, eV) |
|---|---|---|
| **Negative ionization mode (-)** | | |
| 2,4-D-$d_3$ | 244 (32) | 166 (17) |
| Bentazone-$d_6$ | 245 (79) | 132 (29) |
| Bromoxynil-$^{13}C_6$ | 282 (90) | 79 (30) |
| Fenitrothion-$d_3$ | 265 (51) | 153 (22) |
| MCPA-$d_3$ | 204 (38) | 146 (18) |
| Mecoprop-$d_6$ | 218 (38) | 147 (18) |
| **Positive ionization mode (+)** | | |
| Acetamiprid-$d_3$ | 226 (55) | 126 (22) |
| Alachlor-$d_{13}$ | 283 (45) | 251 (11) |
| Atrazine-$d_5$ | 221 (59) | 179 (18) |
| Azinphos-ethyl-$d_{10}$ | 356 (39) | 138 (25) |
| Azinphos methyl-$d_6$ | 324 (43) | 132 (16) |
| Chlorfenvinphos-$d_{10}$ | 369 (58) | 170 (41) |
| Chlortoluron-$d_6$ | 219 (58) | 78 (19) |
| Clothianidin-$d_3$ | 253 (45) | 172 (15) |
| Cyanazine-$d_5$ | 246 (59) | 219 (18) |
| DEA-$d_6$ | 194 (59) | 147 (20) |
| DIA-$d_5$ | 179 (57) | 137 (18) |
| Diazinon-$d_{10}$ | 315 (67) | 170 (23) |
| Dichlorvos-$d_6$ | 227 (69) | 115 (19) |
| Diflufenican-$d_3$ | 398 (84) | 268 (25) |
| Dimethoate-$d_6$ | 236 (44) | 131 (23) |
| Diuron-$d_6$ | 239 (59) | 160 (28) |
| Fenitrothion oxon-$d_6$ | 268 (66) | 222 (19) |
| Fenthion-$d_6$ | 285 (62) | 169 (20) |
| Fenthion oxon-$d_3$ | 266 (69) | 234(17) |
| Fenthion oxon sulfone-$d_3$ | 298 (77) | 218 (20) |
| Fenthion sulfone-$d_6$ | 317 (72) | 131 (17) |
| Fenthion sulfoxide-$d_6$ | 301 (53) | 286 (18) |
| Imidacloprid-$d_5$ | 261 (44) | 214 (19) |
| Irgarol-$d_9$ | 263 (60) | 199 (19) |
| Isoproturon-$d_6$ | 213 (53) | 134 (23) |
| Linuron-$d_6$ | 255 (48) | 185 (18) |
| Malathion-$d_{10}$ | 363 (70) | 237 (18) |
| Methiocarb-$d_3$ | 229 (30) | 169 (10) |
| Metolachlor-$d_{11}$ | 295 (48) | 263 (17) |
| Pendimethalin-$d_5$ | 287 (43) | 213 (12) |
| Propanil-$d_5$ | 223 (59) | 128 (28) |
| Simazine-$d_{10}$ | 212 (63) | 137 (21) |
| Terbuthylazine-$d_5$ | 235 (55) | 179 (18) |
| Terbutryn-$d_5$ | 247 (59) | 191 (19) |
| Thiacloprid-$d_4$ | 257 (60) | 126 (23) |
| Thiamethoxam-$d_3$ | 295 (43) | 214 (13) |
| Thifensulfuron methyl-$d_3$ | 391 (64) | 167 (18) |
| Triallate-$^{13}C_6$ | 310 (55) | 143 (29) |

### *3.2. Method performance*

Figures of merit of the developed methodology are summarized in Tables 4 and 5, and extracted ion chromatograms of the target pesticides in fortified sediment samples are shown in Fig. 3.

As shown in Table 4, the linearity of the method expanded between 0.01 ng/mL to 1000 ng/mL in most cases, up to 500 ng/mL in the case of bromoxynil. Shorter linearity ranges were observed for triallate and pendimethalin, because of their poor sensitivity. Coefficients of determination ($r^2$) were above 0.99 in all cases, except for fenitrothion (0.9847) and pendimethalin (0.9878).

Absolute recoveries, in general good agreement at the three concentration levels, ranged between 21% (fenitrothion oxon) and 99.7% (dimethoate) (average absolute recovery of the figures observed at the three investigated levels). Relative recoveries at any of the investigated concentration levels were always between 76% and 124%. This confirms that the use of surrogate standards (isotopically labelled analogues for 44 pesticides, and isotopically labelled pesticides similar in structure or analytical retention time and absolute recovery for the remaining 6 pesticides fluroxypyr, fenthion oxon sulfoxide, azynphos-methyl oxon, malaoxon, molinate, and quinoxyfen) allows compensating for the potential losses of the analytes during the extraction and clean-up steps and for ionization effects potentially caused by matrix components.

Relative standard deviation (RSD) values obtained after n=6 replicate analysis of fortified samples at the three different concentration levels were always below 20%, except for fluroxypyr (24%) at the low concentration level tested and cyanazine (22%) and desethylatrazine (21%) at the medium concentration level investigated.

**Table 4.** Method performance in terms of linearity ($r^2$), absolute and relative recoveries and repeatability at 50 ng/g, and sensitivity for the target pesticides in sediment.

| Analyte | Linearity ($r^2$) | Recovery[a] | | Repeat.[b] | Sensitivity[c] | |
|---|---|---|---|---|---|---|
| | | Abs. (%) 50 ng/g | Rel. (%) 50 ng/g | RSD (%) 50 ng/g | LOD ng/g d.w. | LODet ng/g d.w. |
| 2,4-D | 0.9901 | 34 | 99 | 20 | 0.59 | 1.97 |
| Acetamiprid | 0.9936 | 91 | 109 | 20 | 0.02 | 0.06 |
| Alachlor | 0.9917 | 49 | 95 | 4 | 0.59 | 1.57 |
| Atrazine | 0.9974 | 31 | 106 | 16 | 0.01 | 0.04 |
| Azinphos ethyl | 0.9901 | 99 | 108 | 7 | 2.19 | 12.4 |
| Azinphos methyl | 0.9916 | 22 | 88 | 10 | 10 | 15 |
| Azinphos methyl oxon[ɣ] | 0.9928 | 81 | 96 | 5 | 0.09 | 0.32 |
| Bentazone | 0.9957 | 32 | 121 | 17 | 0.10 | 0.80 |
| Bromoxynil | 0.9959 | 32 | 86 | 3 | 11.6 | 25.6 |
| Chlorfenvinphos | 0.9945 | 49 | 83 | 4 | 0.57 | 1.9 |
| Chlortoluron | 0.9911 | 61 | 115 | 5 | 0.51 | 1.83 |
| Clothianidin | 0.9916 | 87 | 104 | 15 | 0.18 | 0.62 |
| Cyanazine | 0.9947 | 75 | 83 | 22 | 0.11 | 0.37 |
| DEA | 0.9990 | 48 | 89 | 21 | 0.04 | 0.14 |
| DIA | 0.9966 | 44 | 83 | 5 | 0.11 | 0.38 |
| Diazinon | 0.9949 | 69 | 76 | 3 | 0.03 | 0.10 |
| Dichlorvos | 0.9921 | 42 | 122 | 17 | 50 | 63.1 |
| Diflufenican | 0.9927 | 52 | 99 | 5 | 1.25 | 6.01 |
| Dimethoate | 0.9963 | 100 | 116 | 4 | 0.02 | 0.10 |
| Diuron | 0.9951 | 70 | 89 | 6 | 3.62 | 12.1 |
| Fenitrothion | 0.9847 | 24 | 117 | 13 | 2.18 | 7.26 |
| Fenitrothion oxon | 0.9948 | 22 | 88 | 14 | 48.6 | 62.2 |
| Fenthion | 0.9931 | 21 | 88 | 12 | 50 | 71.5 |
| Fenthion oxon | 0.9971 | 72 | 95 | 4 | 0.02 | 0.06 |
| Fenthion oxon sulfone | 0.9944 | 64 | 112 | 11 | 0.20 | 1.64 |
| Fenthion oxon sulfoxide[x] | 0.9963 | 24 | 85 | 10 | 0.02 | 0.07 |
| Fenthion sulfone | 0.9941 | 23 | 78 | 20 | 50 | 65.4 |
| Fenthion sulfoxide | 0.9911 | 31 | 86 | 20 | 0.20 | 0.68 |
| Fluroxypyr[ɔ] | 0.9912 | 34 | 103 | 7 | 1.27 | 7.79 |
| Imidacloprid | 0.9915 | 74 | 89 | 17 | 0.06 | 0.17 |
| Irgarol | 0.9993 | 26 | 104 | 12 | 0.01 | 0.03 |
| Isoproturon | 0.9981 | 71 | 81 | 5 | 0.61 | 2.04 |
| Linuron | 0.9905 | 50 | 92 | 3 | 10.6 | 31.9 |
| Malaoxon[ᵜ] | 0.9952 | 79 | 82 | 5 | 0.15 | 1.32 |
| Malathion | 0.9928 | 42 | 113 | 6 | 2.23 | 7.46 |
| MCPA | 0.9944 | 21 | 111 | 20 | 1.51 | 5.02 |
| Mecoprop | 0.9913 | 58 | 108 | 16 | 50 | 68.1 |
| Methiocarb | 0.9931 | 98 | 96 | 19 | 1.17 | 3.93 |
| Metolachlor | 0.9980 | 26 | 93 | 5 | 0.02 | 0.23 |
| Molinate[δ] | 0.9930 | 78 | 97 | 6 | 0.35 | 0.98 |
| Pendimethalin | 0.9878 | 23* | 85* | 17* | 100 | 120 |
| Propanil | 0.9910 | 39 | 99 | 16 | 1.85 | 6.17 |
| Quinoxyfen[ɯ] | 0.9974 | 45 | 102 | 7 | 0.33 | 1.31 |
| Simazine | 0.9945 | 34 | 112 | 13 | 0.14 | 0.62 |
| Terbuthylazine | 0.9937 | 35 | 105 | 18 | 0.34 | 1.16 |
| Terbutryn | 0.9909 | 28 | 111 | 11 | 0.02 | 0.05 |
| Thiacloprid | 0.9927 | 106 | 117 | 19 | 0.03 | 0.13 |
| Thiamethoxam | 0.9990 | 76 | 120 | 9 | 0.11 | 0.33 |
| Thifensulfuron methyl | 0.9906 | 48 | 86 | 2 | 0.64 | 2.14 |
| Triallate | 0.9902 | 28* | 85* | 18* | 100 | 121 |

[ɣ] Compound quantified using *fenthion sulfoxide-$d_6$* as surrogate standard
[x] Compound quantified using *thiamethoxam-$d_3$* as surrogate standard
[ɔ] Compound quantified using *mecoprop-$d_3$* as surrogate standard
[ᵜ] Compound quantified using *chlortoluron-$d_6$* as surrogate standard
[δ] Compound quantified using *linuron-$d_6$* as surrogate standard

[ɯ] Compound quantified using *methiocarb-d*$_3$ as surrogate standard

[a] Average absolute recovery: comparison of peak areas obtained in n=6 fortified samples and a methanol standard solution at equivalent concentrations. Average relative recovery: comparison of absolutes recoveries obtained for the analyte and its corresponding surrogate standard (n=6).

[b] Repeatability: relative standard deviation of the relative recoveries observed in n=6 fortified samples.

[c] Sensitivity: limit of detection (LOD) – estimated concentration that would result in a S/N of 3. LODet: limit of determination, minimum concentration at which the analyte can be quantified (LOQ of SRM1) and confirmed (LOD of SRM2). Calculated from sediment fortified samples at 10 ng/g, except for those compounds whose limits of detection were above 10 ng/g.

* Values of recovery and repeatability calculated at 100 ng/g (analytes with LOD>50 ng/g)

**Table 5.** Method performance in terms of absolute and relative recoveries and repeatability at 10 ng/g and 100 ng/g.

| Analyte | Recovery[a] | | | | Repeatability[b] | |
|---|---|---|---|---|---|---|
| | Absolute (%) | | Relative (%) | | RSD (%) | |
| | 10 ng/g | 100 ng/g | 10 ng/g | 100 ng/g | 10 ng/g | 100 ng/g |
| 2,4-D | 31 | 53 | 114 | 115 | 10 | 13 |
| Acetamiprid | 76 | 102 | 116 | 113 | 13 | 19 |
| Alachlor | 46 | 64 | 113 | 94 | 10 | 14 |
| Atrazine | 62 | 31 | 116 | 97 | 2 | 20 |
| Azinphos ethyl | 57 | 73 | 87 | 99 | 12 | 3 |
| Azinphos methyl | 22 | 31 | 82 | 99 | 14 | 6 |
| Azinphos methyl oxon[ɣ] | 102 | 102 | 111 | 104 | 12 | 19 |
| Bentazone | 24 | 31 | 80 | 103 | 8 | 18 |
| Bromoxynil | BLOD | 68 | BLOD | 87 | - | 11 |
| Chlorfenvinphos | 64 | 57 | 107 | 82 | 17 | 12 |
| Chlorotoluron | 83 | 67 | 92 | 88 | 15 | 7 |
| Clothianidin | 72 | 74 | 117 | 87 | 4 | 19 |
| Cyanazine | 99 | 109 | 100 | 111 | 15 | 15 |
| DEA | 39 | 61 | 90 | 92 | 19 | 11 |
| DIA | 64 | 45 | 113.3 | 108 | 2 | 8 |
| Diazinon | 52 | 81 | 114 | 85 | 18 | 14 |
| Dichlorvos | BLOD | 67 | BLOD | 89 | - | 4 |
| Diflufenican | 71 | 69 | 92 | 120 | 13 | 6 |
| Dimethoate | 105 | 94 | 120 | 102 | 2 | 19 |
| Diuron | 55 | 57 | 91 | 106 | 12 | 4 |
| Fenitrothion | 30 | 20 | 81 | 86 | 8 | 16 |
| Fenitrothion oxon | BLOD | 20 | BLOD | 76 | - | 18 |
| Fenthion | BLOD | 26 | BLOD | 81 | - | 17 |
| Fenthion oxon | 68 | 54 | 119 | 119 | 6 | 7 |
| Fenthion oxon sulfone | 99 | 89 | 121 | 103 | 1 | 18 |
| Fenthion oxon sulfoxide[x] | 55 | 51 | 89 | 105 | 1 | 4 |
| Fenthion sulfone | BLOD | 26 | BLOD | 90 | - | 16 |
| Fenthion sulfoxide | 33 | 22 | 87 | 93 | 10 | 13 |
| Fluroxypyr[ↄ] | 27 | 46 | 121 | 88 | 24 | 19 |
| Imidacloprid | 84 | 88 | 98 | 98 | 14 | 15 |
| Irgarol | 39 | 24 | 93 | 88 | 7 | 4 |
| Isoproturon | 94 | 81 | 118 | 87 | 2 | 13 |
| Linuron | BLOD | 50 | BLOD | 102 | - | 12 |
| Malaoxon[*] | 102 | 75 | 111 | 78 | 1 | 14 |
| Malathion | 53 | 33 | 78 | 92 | 4 | 19 |
| MCPA | 26 | 34 | 81 | 97 | 6 | 5 |
| Mecoprop | BLOD | 42 | BLOD | 124 | - | 16 |
| Methiocarb | 110 | 88 | 87 | 117 | 17 | 16 |
| Metolachlor | 21 | 25 | 107 | 86 | 17 | 4 |
| Molinate[δ] | 88 | 77 | 88 | 84 | 8 | 4 |
| Pendimethalin | BLOD | 23 | BLOD | 85 | - | 17 |
| Propanil | BLOD | 41 | BLOD | 88 | - | 17 |
| Quinoxyfen[ɯ] | 46 | 38 | 82 | 110 | 3 | 7 |
| Simazine | 57 | 62 | 96 | 85 | 10 | 18 |
| Terbuthylazine | 54 | 40 | 117 | 92 | 15 | 11 |
| Terbutryn | 20 | 46 | 106 | 99 | 5 | 15 |
| Thiacloprid | 87 | 67 | 104 | 87 | 12 | 8 |
| Thiamethoxam | 96 | 85 | 120 | 96 | 12 | 12 |
| Thifensulfuron methyl | 68 | 56 | 120 | 86 | 5 | 13 |
| Triallate | BLOD | 28 | BLOD | 85 | - | 18 |

[ɣ] *fenthion sulfoxide-$d_6$* used as surrogate standard; [x] *thiamethoxam-$d_3$* used as surrogate standard; [ↄ] *mecoprop-$d_3$* used as surrogate standard; [*] *chlortoluron-$d_6$* used as surrogate standard; [δ] *linuron-$d_6$* used as surrogate standard [ɯ] *methiocarb-$d_3$* used as surrogate standard. BLOD: below limit of detection

[a] Average absolute recovery: comparison of peak areas obtained in n=6 fortified samples and a methanol standard solution at equivalent concentrations. Average relative recovery: comparison of absolutes recoveries obtained for the analyte and its corresponding surrogate standard (n=6). [b] Repeatability: relative standard deviation of the relative recoveries observed in n=6 fortified samples.

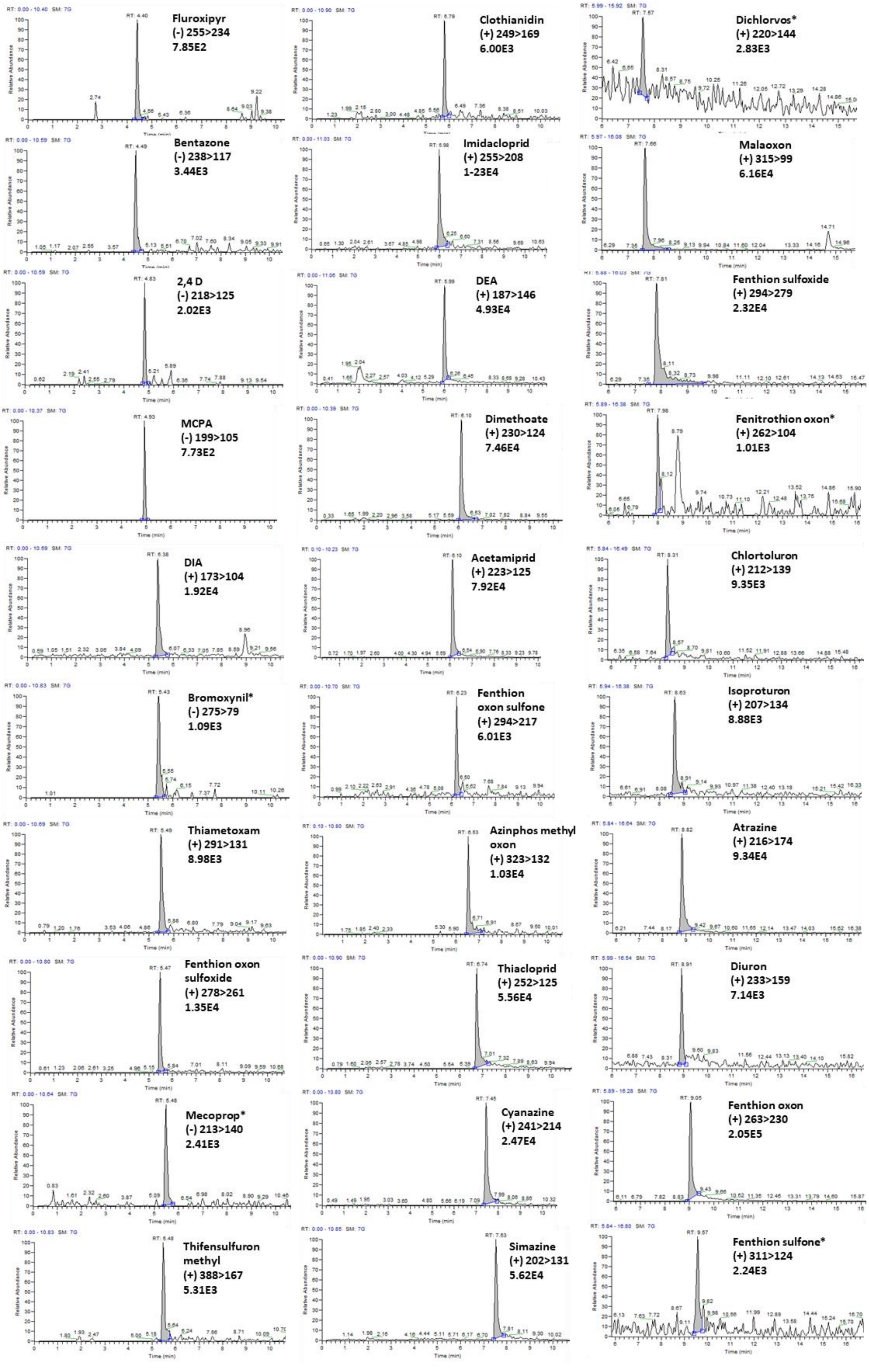

**Figure 3.** Extracted ion chromatograms of the target pesticides after PLE-SPE-LC-MS/MS analysis of a sediment sample fortified at a concentration of 10 ng/g (*50 ng/g for those compounds with LOD above 10 ng/g; **100 ng/g for those compounds with LOD above 50 ng/g).

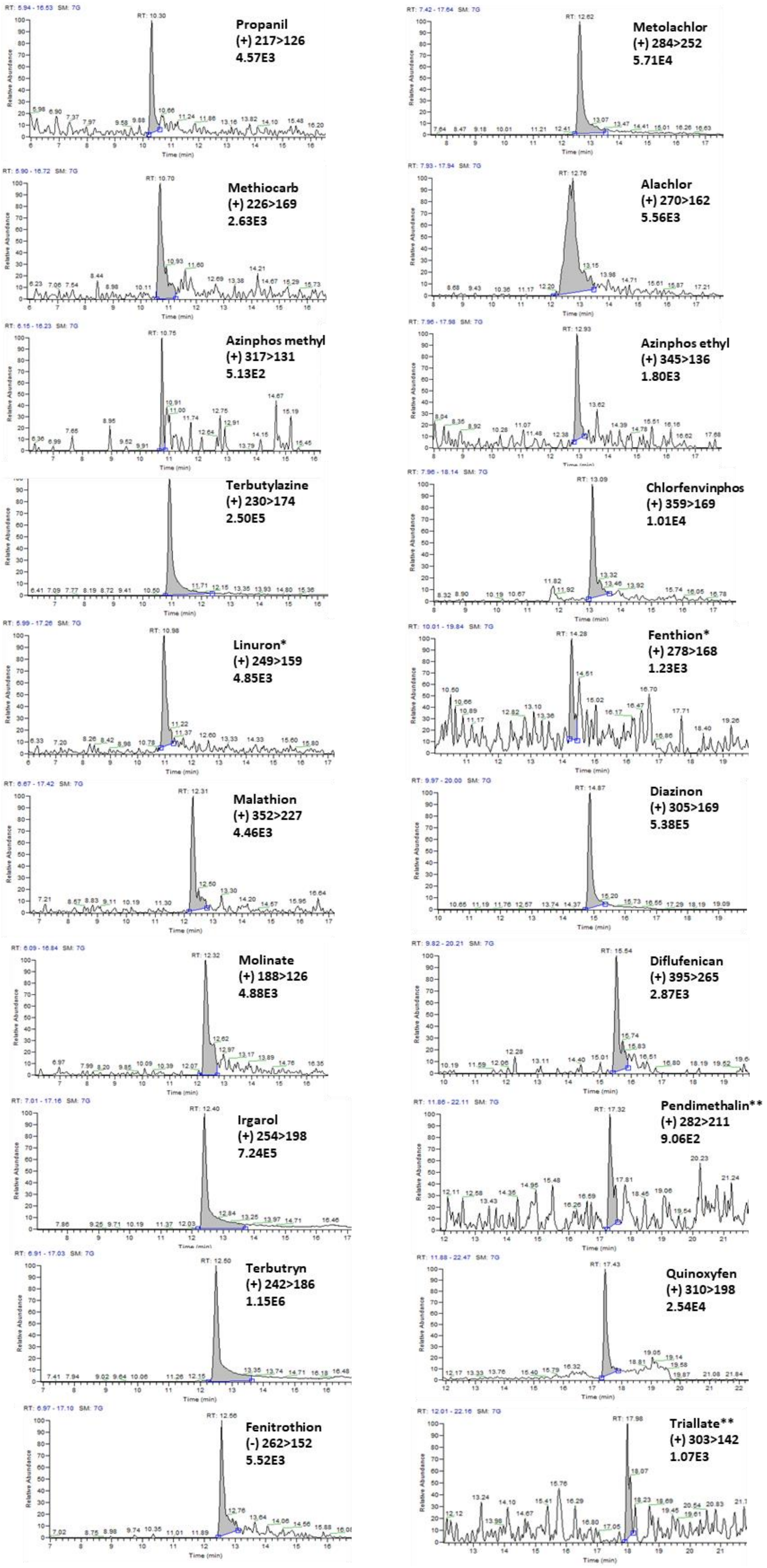

**Figure 3.** (continued).

Results were good in terms of sensitivity for most of the compounds, with LODs between 0.01 and 4 ng/g d.w. and LODets between 0.03 and 12.4 ng/g d.w. for 80% of the investigated compounds. In the absence of legislation establishing maximum pesticide residues in sediments, the results obtained are satisfactory when compared to the value of 50 ng/g set by the European Commission as the desired LOQ for the analysis of pesticide residues in soil [38].

As for the priority substances considered, the method allows determining them in sediments at the low ng/g level, that is at levels similar to the EQS established for these compounds in water (2013/39/EU) (i.e., alachlor (1.6 ng/g d.w. *vs.* 0.3 µg/L), atrazine (0.04 ng/g d.w. *vs.* 0.6 µg/L), chlorfenvinphos (1.9 ng/g d.w. *vs.* 0.1 µg/L), diuron (12 ng/g d.w. *vs.* 0.2 µg/L), irgarol (0.03 ng/g d.w. *vs.* 0.0025 µg/L), isoproturon (2.0 ng/g d.w. *vs.* 0.3 µg/L), simazine (0.6 ng/g d.w. *vs.* 1 µg/L), quinoxyfen (1.3 ng/g d.w. *vs.* 0.15 µg/L), and terbutryn (0.05 ng/g d.w. *vs.* 0.065 µg/L)), except in the case of dichlorvos (63.1 ng/g d.w. *vs.* 0.0006 µg/L). Besides dichlorvos, the worst performance of the method in terms of sensitivity was observed for bromoxinyl (25.6 ng/g d.w), linuron (31.9 ng/g d.w), fenitrothion oxon (62.2 ng/g d.w), fenthion sulfone (65.4 ng/g d.w), mecoprop (68.1 ng/g d.w), fenthion (71.6 ng/g d.w), pendimethalin (120 ng/g d.w), and triallate (121 ng/g d.w). As for the Watch List pesticides, the presented methodology works well to detect neonicotinoids (0.02-0.18 ng/g d.w. *vs.* 0.008 µg/L) and may not be sensitive enough in the case of methiocarb (1.2 ng/g d.w. *vs.* 0.002 µg/L), if maximum acceptable LODs suggested by the EU Decision 2018/840 for these compounds are considered. The effect of matrix components on the LC-MS/MS analysis of the target compounds is summarized in Fig. 4. As shown, the matrix components remaining after extraction did not have a strong effect on the ionization of the analytes, being the signal suppressed or enhanced by less than 10% in all cases.

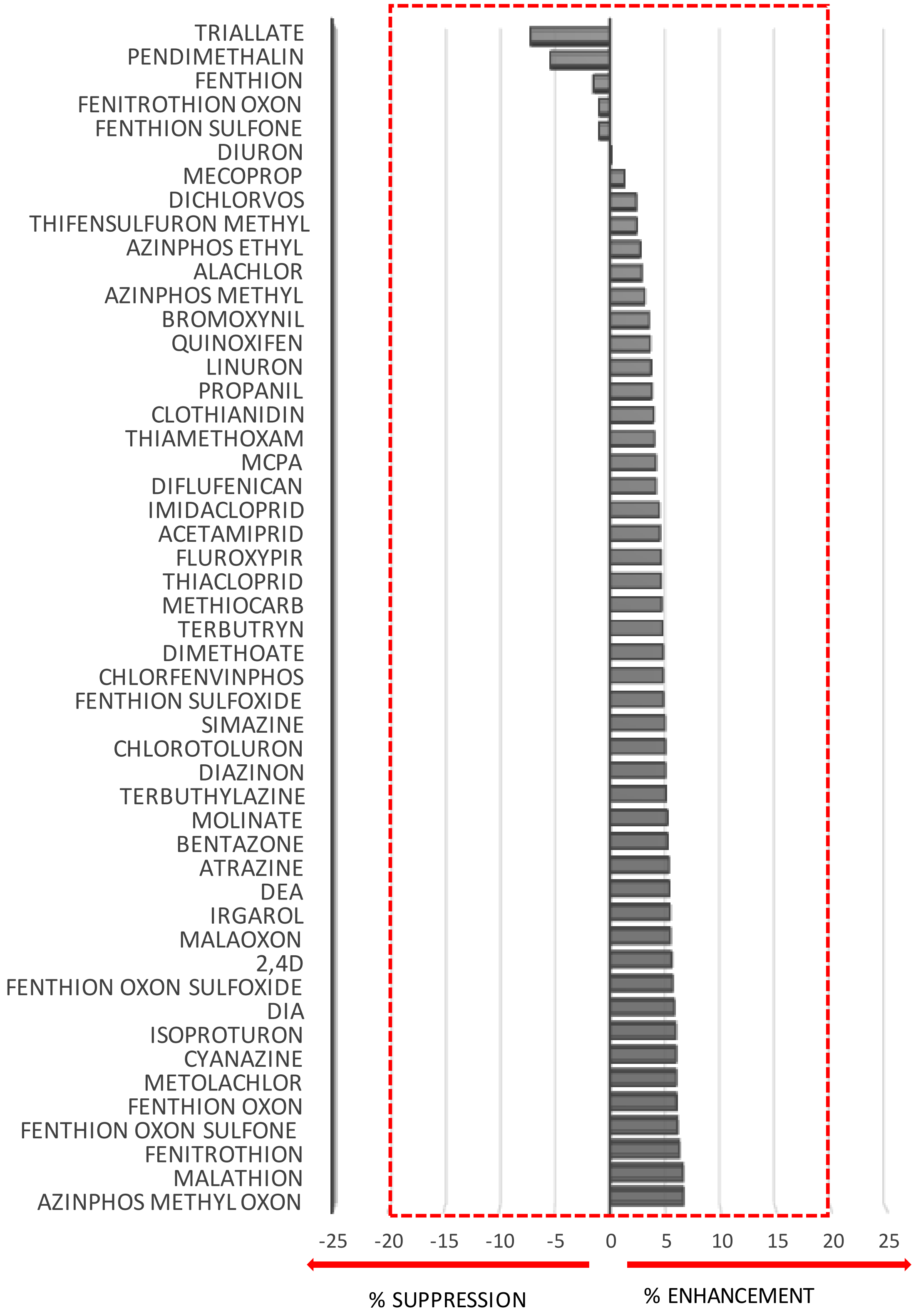

**Figure 4.** Matrix effects calculated for the analysis of the target pesticides in sediment samples.

The performance of this methodology was compared with that of the analytical LC-MS-based methods published in the peer-reviewed literature for the same purpose since 2015 (see Table 6) [17, 18, 20, 14, 15, 21]. The present methodology, together with the methods reported by Farré et al. [21], Masiá et al. [18] and Massei et al. [24], are the only multi-residue analytical approaches currently available for the simultaneous determination of a large number (>50) of pesticides belonging to different chemical classes. Furthermore, the target analytes included in the present method overlapped only by 50-60% with the analyte lists considered in the previously published multi-residue methods. This is the first time indeed that an analytical method is validated for the analysis of azinphos-methyl oxon, fenitrothion oxon, and malaoxon, in sediment samples. Malaoxon was included in the list of target pesticides analyzed by Kalogridi et al. [17] in sediments (253 pesticides in total); however, figures of merit of the method for this and many other pesticides were not provided. Indeed, method performance was partially proven for only 10% of the target pesticides, which undermines results reliability [17].

All analytical approaches recently published to determine pesticides in sediments use matrix-matched calibration curves for quantification, except the one reported by Massei et al [24]. This and the present method use isotopic labelled standards (ILS) for analyte quantification. While Massei et al. use 15 ILS to correct the response of 118 compounds, our method is the first one that uses isotopically labelled analogs for 85% of the 50 target pesticides. In the isotope dilution method, the ILS are added at a known concentration to the sample at the beginning of the extraction process and the analyte response is normalized to that of the corresponding ILS. Contrary to the matrix-matched calibration, which is also highly labor intensive, the isotope dilution method allows, as aforementioned, to correct potential analyte losses during sample preparation and matrix effects and signal drift during the MS analysis. This ensures extremely good precision and accuracy (see relative recoveries in Tables 2 and 3) and hence reliability of the results.

**Table 6. LC-MS-based methodologies published in the peer-reviewed literature since 2015 for the simultaneous determination of medium to highly polar pesticides in sediments.**

| Number of pesticides | Analyte overlap | Extraction step | | Extract purification[c] | Quantification method[f] | Accuracy (Analyte recovery, %) | Precision, (RSD %) | Sensitivity (LOQ, ng/g) | Matrix effects | Reference |
|---|---|---|---|---|---|---|---|---|---|---|
| | | Technique[a] | Solvent[b] | | | | | | | |
| 50 | | PLE | Acet:DCM (1:1, 1% FA *v/v*) | SPE | isotope dilution[d] (ILS for 85% of compounds) | 21-99 absolute rec.– 76-124 relative rec. | <20 | 0.03-12.4 (80% of analytes); 23-121 for the others | ± 10 | This study |
| 12 | 6 | USE | MeOH:DCM (1:1) | none | matrix-matched | 12-125 | <26 | 3– 10 | ± 30 | [14] |
| 17 | 7 | MSPD | 20 mL EtAc + 5 mL ACN | none | matrix-matched | 60.9-99.7 | <0.73 | 0.23–4.26 | not provided | [15] |
| 253[e] | 31 | MAE | ACN: Hexane | 8 g of $Na_2SO_4$ | not provided | 67-123 [e] | not provided | 0.009-0.072 [e] | not provided | [17] |
| 50 | 27 | QuEChERS | ACN | dSPE | matrix-matched | 39-120 | <25 | 0.1-15 | 0-250 | [18] |
| 18 | 7 | QuEChERS | ACN (0.1% acetic acid) | dSPE | matrix-matched | 70.8-106 | <21 | 0.8-13 | ± 20 | [20] |
| 54 | 27 | QuEChERS | 7.5 mL $H_2O$ + 10 mL ACN | none | matrix-matched | 59-113 | <20 | 0.1-2 ng/mL (LOD) | ± 50 | [21] |
| 118 | 33 | PLE | EtAc:Acet (1:1)+ Acet:FA 1% + MeOH:$H_2O$ (9:1) | NPChrom or solvent exchange | isotope dilution (15 ILS) | 11-123 (overall) | <20 (overall) | 0.016-12.8 | not provided | [24] |

[a] *MAE:microwave assisted extraction, MSPD: matrix solid phase dispersion extraction, PLE: pressurized liquid extraction, USE: ultrasonic solvent extraction*

[b] *ACN: acetonitrile; DCM: dichloromethane; EtAC: ethyl acetate; FA: formic acid; MeOH: Methanol*

[c] *SPE: solid phase extraction, dSPE: dispersive solid phase extraction; NPChrom: normal phase chromatography*

[d] *isotopically labelled analogues are available for 85% of the target compounds*

[e] *data regarding method performance is provided for only 25 pesticides*

[f] *ILS: isotopic labelled standards*

As compared to the other analytical methods, the present one, despite being more labor demanding and time consuming than QuEChERs approaches, is highly effective in terms of removing matrix interferences.

Method sensitivity, with LOQs between 0.03 ng/g and 12.4 ng/g for most compounds, is similar to that reported in other LC-MS-based methods (see Table 6). The present method provides even lower LOQs for few compounds, namely acetamiprid, atrazine, fenthion oxon, fenthion oxon sulfoxide, irgarol and terbutryn (<0.1 ng/g d.w) as compared to previously published methods. Exceptionally low LOQs (0.1 and 0.01 ng/g) were reported by Kalogridi et al. [17]; however, as previously indicated, the method used was not rigorously validated and few details were provided on LOQ calculation in this study.

### *3.3. Application to real samples*

The validated method was applied to the analysis of real sediment samples with a double objective: i) to test its applicability and effectiveness and ii) to assess the presence of the target pesticides in the lower Llobregat River basin (NE Spain). Only 5 out of the 50 compounds investigated, namely, terbutryn, dichlorvos, terbuthylazine, diazinon, and irgarol, were present in the seven sediment samples investigated (see Fig. 4). The most relevant compound in terms of abundance and ubiquity was the herbicide terbutryn that was found in all samples and showed a maximum concentration of 200 ng/g d.w. (Pt. 1). Terbutryn is an herbicide used as a pre-emergent and post-emergent control agent for grasses and broadleaf weeds in various cultivations (e.g. wheat, barley, sunflower, potatoes) and as an aquatic herbicide for the control of algae in water courses, reservoirs, and fish ponds. Previous studies already showed the presence of terbutryn at trace levels (5 ng/g d.w.) in sediment samples collected in the Llobregat River basin in 2011 [39], and in sediment samples from other river basins in Spain, like the Ebro River basin where it was found at a maximum concentration of 22 ng/g d.w. [40]. The results obtained in the present study reveal a significant increase of this pesticide in the Llobregat River sediment, in spite of the fact that the concentration of pollutants in this matrix may strongly depend on the season and the climatic conditions in which the samples are collected. Since terbutryn is one of the EU priority substances [8] with a maximum EQS of 34 ng/L in water, the evolution of its concentrations in the sediments of this basin should be monitored.

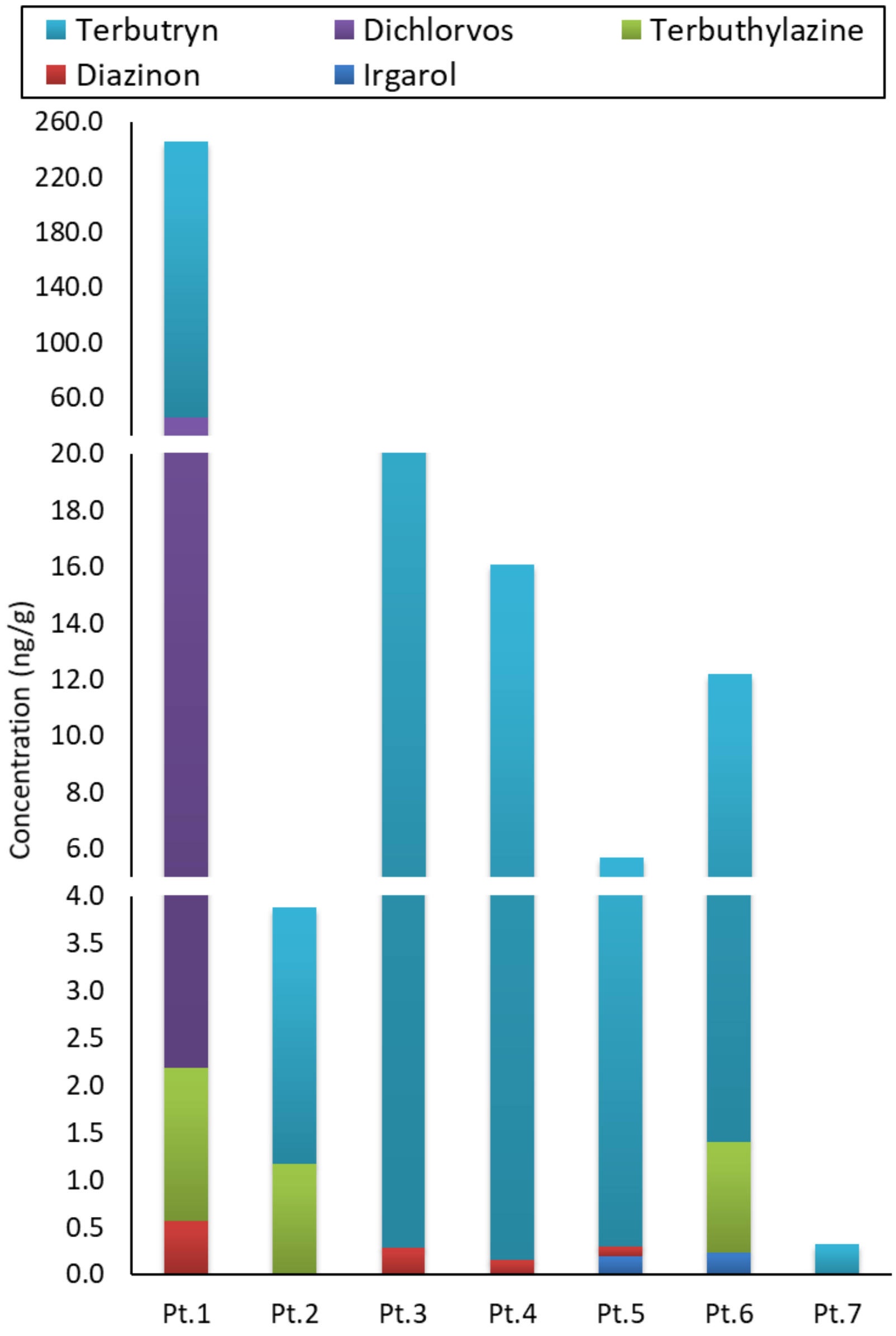

**Figure 4.** Concentrations of pesticides found in sediment samples of the Llobregat River basin.

The second most abundant pesticide detected in the lower Llobregat River basin sediments was the insecticide dichlorvos, found at a concentration of 44 ng/g d.w. in one sample (Pt.1). This insecticide has various applications: insect control in the workplace and at home, in food-storage areas, greenhouses and vegetable crops; and parasite control in dogs, livestock and humans. Dichlorvos is also considered as a priority substance in the EU, with a very low EQS (maximum allowable concentration of 0.07 ng/L in water). Although dichlorvos residues in sediments are not regulated, the concentration found in the present study exceeds by far the EQS in water. This insecticide was also reported to be present in sediments from the River Wuchuan in Southeast China (0.23 ng/g) [41] and in sediments from and around a highly eutrophic lake in Eastern China (up to 23.3 ng/g d.w.) [42].

Irgarol, diazinon, and terbuthylazine were all found at trace levels (up to 0.2, 0.6, and 1.6 ng/g, respectively) in the investigated sediment samples. To the author's knowledge, this is the first time that irgarol, an algaecide used in antifouling paints for boats, is detected in Llobregat River sediments. However, since it was not previously investigated in this area, its previous presence cannot be discarded. The presence of diazinon and terbuthylazine in Llobregat River sediments has been studied in a few occasions, and trace levels of these pesticides were always found. In the case of diazinon, with past levels up to 4.6 ng/g d.w. [29, 39, 27], its presence (up to 0.6 ng/g in the present work) could be attributed to the proximity of the sampling locations to important urban zones (Barcelona and its metropolitan area), where diazinon could be extensively used for insect control. Terbuthylazine is an endemic pesticide used to control a broad spectrum of weeds, frequently found in the water of the Llobregat River basin [43, 44] while in sediment the only previous study conducted in this area [29] reported low levels (up to 1.4 ng/g d.w.) in the same range of those detected in the present study (up to 1.6 ng/g d.w.).

### *3.4. Environmental fate and ecotoxicity risk assessment*

To better understand the fate of pesticides in the aquatic environment and the potential impact that they may pose to aquatic organisms, their physical-chemical properties need to be considered as they determine their mobility, adsorption and bioaccumulation capacity , and degradation, and hence their environmental concentrations. One of the most useful properties for assessing pesticide accumulation potential in organisms and adsorption potential onto particles is the octanol-water partition coefficient ($K_{ow}$). Compounds with high Log $K_{ow}$ values (>3) have low affinity for water and are more readily sorbed onto particles. This process affects pesticide mobility in environmental systems [45]. Mobility of a pesticide is high if its solubility is high and its organic carbon-water partition coefficient ($K_{oc}$) is low. These properties, together with the degradation potential, expressed by the half-life ($DT_{50}$), determine the potential of a pesticide to move through soils and leach into groundwater (also assessed through the Groundwater Ubiquity Score (GUS) index [46]), or move by runoff to surface water bodies [47].

According to this, all pesticides detected in the sediment samples analyzed, except dichlorvos, present a Log $K_{ow}$ > 3 (see Table 1) and can therefore potentially bioaccumulate in aquatic organisms. Many other of the pesticides investigated present also a high potential for accumulation (Table 1), but they have not been detected, despite that many of them (e.g. chlortoluron, diflufenican, isoproturon, linuron, and pendimethalin) have been largely applied in Catalonia. This could be explained by a distinct pattern of pesticide application in the investigated area compared to other Catalan regions. On the contrary, the non-detection of other pesticides extensively used also in the investigated territory (i.e., 2,4 D, bentazone, bromoxynil, fluroxypyr, and MCPA) in the sediment samples analyzed could be attributed to their low Log $K_{ow}$ and high solubility.

Of all pesticides detected in the present study, terbuthylazine is the only one that exhibits a GUS index above 2.8 and is therefore likely to leach into groundwater [48]. Meanwhile, terbutryn and irgarol are the pesticides less likely to desorb from the sediments due to their high adsorption potential (with Koc values of 2432 mL/g and 1569 mL/g, respectively) and low water solubility (25 mg/L in the case of terbutryn and 7 mg/L in the case of irgarol). Notwithstanding this, events such as changes in water flow or heavy rainfall and human activities could result in their eventual release in the water column.

Overall, the accumulation of pesticides in sediments can pose a direct toxicological risk for organisms living and feeding on river sediments due to chronic exposure. The potential ecotoxicity risk that the pesticides found may pose to aquatic organisms was assessed using the hazard quotient (HQ) approach. This approach compares the measured environmental concentrations (MEC) with predicted no-effect concentrations (PNEC) at which no toxic effects are expected. In this study, we used the maximum concentration measured for each pesticide in the investigated samples as MEC and the PNEC values in sediments ($PNEC_{sed}$) were extracted from the NORMAN Ecotoxicology Database. These values are derived from the corresponding PNECs of the pesticides in water (which are predicted by QSAR models or obtained experimentally) after applying an equilibrium partitioning approach [49]. According to this approach, pesticides showing HQ values < 1 are not considered hazardous for aquatic ecosystems, pesticides exhibiting HQ values between 1 and 10 are considered as potentially hazardous, and pesticides with HQ values >10 are considered as the most hazardous to aquatic organisms.

HQ values calculated for the five pesticides detected in the sediments of the Llobregat River are shown in Table 7. In all cases, HQ values above 10 were obtained indicating high risk for aquatic organisms. This is due to the low PNEC values of the pesticides detected (<0.1 ng/g), that result in HQ>10 even at low environmental concentrations, as in the case of terbuthylazine. However, the extraordinarily high HQ values obtained for terbutryn (2000) and dichlorvos (45250) are also explained by the relatively high levels of these pesticides measured in the sediments analyzed. These findings are of concern especially in the case of terbutryn because of its widespread use in the investigated area.

**Table 7.** Hazard quotient (HQ) values calculated for the pesticides detected in Llobregat River sediments.

| Analyte | MEC[a] (ng/g) | $PNEC_{sed}$[b] (µg/Kg) | HQ |
|---|---|---|---|
| Irgarol | 0.2 | 0.005 | 42 |
| Diazinon | 0.6 | 0.016 | 35 |
| Terbuthylazine | 1.6 | 0.096 | 17 |
| Dichlorvos | 44 | 0.001 | 45250 |
| Terbutryn | 200 | 0.1 | 2000 |

[a] MEC: maximum environmental concentration measured

[b] PNEC values extracted from https://www.norman-network.com/nds/ecotox/[49]

## *4.* Conclusions

This study presents an analytical methodology based on PLE extraction, SPE clean-up, and LC-HESI-MS/MS analysis for the determination of 50 moderately polar pesticides in sediment samples. The method was validated at three different concentration levels. Validation results indicate that the method is satisfactory in terms of sensitivity (with LODs below 4 ng/g d.w. and LODets below 12.4 for 40 of the 50 compounds), accuracy (with relative recoveries between 76% and 124% in all cases), and precision (with RSD always below 20%). Besides, it is very effective in removing matrix interferences, which justifies the use of a long sample treatment approach that gets simplified and less labor intensive with the automation of the PLE process. Despite the fact that the described methodology presents similar sensitivity than previously published LC-MS-based methods, the use of an isotope dilution approach for quantification is a clear advantage over them; because results obtained in this way are highly reliable. This is the first time that azinphos-methyl oxon, fenitrothion oxon, and malaoxon are investigated in sediment samples.

The presented methodology allows evaluating a wide spectrum of pesticides belonging to many different classes and used for many different purposes, and due to its multi-residue character, the list of target analytes could be extended with pesticides that belong to those chemical classes covered in the current validated list. In this regard, evaluation of method performance of newly included pesticides is always recommended.

The application of the method to sediment samples collected at the lower Llobregat River basin revealed the presence of 5 pesticides, namely, terbutryn, dichlorvos, terbuthylazine, diazinon, and irgarol. All of them, but dichlorvos, present a high potential to sorb onto particles (Log Kow > 3). To the author's knowledge, this is the first evidence of the presence of irgarol in sediments of the Llobregat River. According to the pesticides physical-chemical properties, the presence of terbuthylazine in sediments is of concern, since it presents a high leaching potential into groundwater (GUS of 3.07) if released from the sediment. This pesticide has been continuously detected in the Llobregat River water since the 90s, which may be attributed to its extended use for both urban and agricultural purposes. Likewise, terbutryn (low water solubility and high $K_{oc}$), which was the most ubiquitous and abundant compound, raises concern since sediments represent a diffuse source of this pollutant into the water. According to the HQ approach, all 5 pesticides pose an environmental risk for aquatic organisms living and feeding on sediments. The risk predicted is explained mainly by their low PNEC values, and in the case of terbutryn and dichlorvos also by the high levels detected. However, additional studies must be conducted to better understand the toxicity of pesticide-polluted sediments on aquatic communities.

## Acknowledgments

This work has received funding from the Government of Catalonia (2017 SGR 01404), the Spanish State Research Agency (AEI) and the European Regional Development Fund (ERDF) through the project BECAS (grant number CTM2016-75587-C2-2-R), and the European Union's Horizon 2020 Research and Innovation Programme under grant agreement No. 727450. This manuscript only reflects the authors' views and the Commission is not responsible for any use that may be made of the information it contains.

**Compliance with ethical standards:**

This study did not involve any human participants or animals

**Conflicts of interests:** The authors declare that they have no conflicts of interest